\documentstyle[12pt,world_sci]{article}
\begin{document}
\hbox{OITS-573}

\title{{\bf TEST OF CP VIOLATION IN NON-LEPTONIC HYPERON DECAYS}}
\author{XIAO-GANG HE\thanks{Talk presented at the Beyond the Standard Model IV,
Lake Tahoe,
California, December 13 - 18, 1994}\\
{\em Institute of Theoretical Science, University of Oregon\\
Eugene, OR 97403-5203, USA}}

\maketitle
\setlength{\baselineskip}{2.6ex}

\begin{center}
\parbox{13.0cm}
{\begin{center} ABSTRACT \end{center}
{\small \hspace*{0.3cm} In this talk I discuss CP violation in hyperon decays
in  Left-Right
 symmetric models. I show that the asymmetry in polarization in $\Lambda
\rightarrow p\pi^-$ can
be as large as $6\times 10^{-4}$ in these models, which is an order of
magnitude larger than the
Standard Model prediction.}}
\end{center}

\section{Introduction}
Non-leptonic hyperon decay, $\Lambda\rightarrow p\pi^-$, is an interesting
process to test CP
 conservation
outside the neutral Kaon system\cite{1,2,3,4}.The decay amplitude can be
written as
\begin{eqnarray}
M(\Lambda \rightarrow p\pi^-) =S + P\vec \sigma\cdot \vec q\;,
\end{eqnarray}
where $\vec q$ is the momentum direction vector of the
pion. One particularly interesting CP violating observable
is the asymmetry in polarization,
\begin{eqnarray}
A(\Lambda)  = {\alpha+\bar \alpha\over \alpha-\bar \alpha}\;,
\end{eqnarray}
where $\alpha = 2 Re(S^*P)/(|S|^2 +|P|^2)$, and $\bar \alpha$ is the
corresponding quantity for $\bar \Lambda$ decays. A non-zero $A(\Lambda)$
signals CP
violation. To a very good approximation,
\begin{eqnarray}
A(\Lambda) =
-\mbox{tan}(\delta_{11}-\delta_1)\mbox{sin}(\phi^p_1-\phi^s_1)\;,
\end{eqnarray}
where $\delta_{1} = 6^0$, $\delta_{11} = -1.1^0$ are the strong rescattering
phases\cite{5},
 $\phi^{s}_1$ and $\phi^p_1$ are the weak CP violating phases for the S-wave
and P-wave amplitudes
with $I=1/2$, respectively. To calculate the weak phases $\phi^{s,p}$, I will
take the approach to
use experimentally determined decay amplitudes as CP conserving ones and
calculate theoretically the CP violating ones.

A new experiment
E871 at Fermilab will measure $\alpha_\Lambda \alpha_\Xi$ in the
decay $\Xi^- \rightarrow \Lambda \pi^-\rightarrow p\pi^-\pi^-$, and also
similar measurement for
anti-$\Xi$ decay\cite{6}. CP asymmetry in these
decays will be measured. The CP asymmetry in polarization in this case is
dominated by
$A(\Lambda)$\cite{7}. The expected sensitivity for $A(\Lambda)$ is $10^{-4}$
and eventually will
reach $10^{-5}$. In the SM $A(\Lambda)$ is predicted to be in the range
 $-(0.5 \sim 0.1)\times 10^{-4}$\cite{3,4}.
This prediction will not be tested at the initial stage of the E871 experiment.
It is exptremely interesting to see if $A(\Lambda)$ can be larger in
extensions of the SM and can be tested by the E871 experiment.
In this talk I will show that in Left-Right symmetric models $A(\Lambda)$
can be as large as $6\times 10^{-4}$ and will be tested soon\cite{8}.

\section{$A(\Lambda)$ in Left-Right Symmetric Models}

Left-Right symmetric extensions of the SM is based on the gauge group
$SU(3)_C\times
SU(2)_L\times SU(2)_R\times U(1)_{B-L}$\cite{9}. In general there will be
mixing between $W_L$ of the $SU(2)_L$ charged gauge boson and $W_R$ of the
$SU(2)_R$.
The effective Hamiltonian $H_{eff}$ for non-leptonic hyperon decays by
exchanging gauge bosons
up to one
loop level contains different contributions
\begin{eqnarray}
H_{eff} = H_{SM} + H_{R} + H_{LR}\;.
\end{eqnarray}
In the zero mixing limit, $H_{SM}$ reduces to the SM contribution.
$H_R$ is due to $W_R$ exchange. $H_{LR}$ is due to $W_L$ and $W_R$
mixing and therefore should be proportional to the mixing angle $ \xi$.  It is
given by\cite{8}
\begin{eqnarray}
H_{LR} &=& {G_F\over \sqrt{2}}\bar \xi\{
V^*_{Lud}V_{Rus}(O_+^{LR}\eta_+ - O_-^{LR}\eta_-)
+V^*_{Rud}V_{Lus}(O_+^{RL}\eta_+ - O_-^{RL}\eta_-)\nonumber\\
&+&\sum_i \tilde G(x_i) {g_s\over 16 \pi^2} m_i\eta_g
G^{a\mu\nu} \bar d \sigma_{\mu\nu}\lambda^a
[V_{Rid}^*V_{Lis} (1-\gamma_5) + V_{Lid}^*V_{Ris}(1+\gamma_5)]s\}\;,
\end{eqnarray}
where $\eta_+ =
(\alpha_s(1GeV)/\alpha_s(m_{c}))^{-3/27}(\alpha_s(m_c)/\alpha_s(m_{b}))^{-3/25}(\alpha_s(m_b)/
\alpha_s(m_{W_1}))^{-3/23}$, $\eta_- = \eta_+^{-8}$,
$\eta_g =\eta_+^{14/3}$, $\bar \xi = \xi g_R/g_L$, $g_{L,R}$ are the
$SU(2)_{L,R}$ gauge couplings,
$V_{L,R}$ are the KM
mixing matrices, and
\begin{eqnarray}
\tilde G(x) &=& -{3x\over 2(1-x)^3}\mbox{ln}x - {4+x+x^2\over
4(1-x)^2}\;,\nonumber\\
O^{LR}_+ &=& \bar d \gamma_\mu Lu\bar u \gamma^\mu Rs
+{2\over 3} \bar d R s \bar u L u\;,\;\;
O^{LR}_- = {2\over 3} \bar d R s \bar u L u\;.
\end{eqnarray}
Here $R(L) = 1\pm \gamma_5$. $O^{RL}_{\pm}$ are obtained by exchanging $R$ and
$L$.

$H_{SM}$ contribution to $A(\Lambda)$ is in the range of $-(0.5 - 0.1)\times
10^{-4}$, and $H_R$
contribution is smaller\cite{3,4}. Possible large contribution can come from
$H_{LR}$.
Using factorization approximation, the $O_{\pm}^{ij}$ operator contribution
$A_W(\Lambda)$ is
estimated to be\cite{8}
\begin{eqnarray}
A_W(\Lambda) &=& 1.73 \{\xi^u_-  - 0.06 \xi^u_+ \}\;,
\end{eqnarray}
where
$\xi^i_\pm = \bar \xi  Im(V_{Lid}^*V_{Ris} \pm V_{Rid}^*V_{Lis})$.
Using the matrix element evaluated in Ref.[3], the gluon dipole contribution
$A_G(\Lambda)$
(terms proportional to $\tilde G(x)$) is found to be\cite{8}
\begin{eqnarray}
 \;\;A_G(\Lambda) = 0.21 \sum_i
\tilde G(x_i){m_i\over GeV}
\{ \xi^i_- + 1.17 \xi^i_+ \}\;.
\end{eqnarray}
The S-wave and P-wave
contributions are proportional to $\xi^i_-$ and $\xi^i_+$, respectively.

\section{Discussions}
There are contraints on the allowed value for $A(\Lambda)$ because $\bar \xi$
and $\xi^u_-$ are
both
constrained. From CP conserving hyperon decays, $\bar \xi$ is found to be less
than $4\times
10^{-3}$\cite{10}.
Experimental limit of $|\epsilon'/\epsilon| < 3 \times 10^{-3}$ implies
$|\xi^u_-|< 2\times
10^{-6}$\cite{8,11}. The S-wave contribution to
$A_W(\Lambda)$ is constrained to be less than $4\times 10^{-6}$. If $\xi^u_-$
is small by cancellation,
the P-wave contribution is approximately given by
\begin{eqnarray}
A_W(\Lambda) &=& -0.2\bar\xi Im(V_{Rud}^*V_{Lus})\;.
\end{eqnarray}
$A_W(\Lambda)$ can be as large as $10^{-4}$ if
$Im(V_{Rud}^*V_{Lus}) >0.1$.

The magnitude for $A_G(\Lambda)$ is also constrained for the same reasons
discussed above.
 The S-wave
contribution to $A_G(\Lambda)$ is
found to be less than $1.3\times 10^{-4}$ from experimental limit on
$\epsilon'/\epsilon$. However
the P-wave contribution is not directly contrained and may be larger. Assuming
$\mbox{Im}
(V^*_{Lid}V_{Ris}) = \mbox{Im}(V^*_{Rid}V_{Lis})$ and using $m_t = 176$ GeV,
one finds that
$A_G(\Lambda)$ can be as large as $6\times 10^{-4}$.
Such a large $A(\Lambda)$ will be probed by the E871 experiment.

I thank Professors D. Chang and S. Pakvasa for collaborations for the work
reported here.

\bibliographystyle{unsrt}

\end{document}